\documentclass{article}

\usepackage{arxiv}

\usepackage[utf8]{inputenc} 
\usepackage[T1]{fontenc} 
\usepackage{hyperref, amsmath, amsfonts, microtype, graphicx, physics}
\usepackage[capitalise]{cleveref}

\title{The role of noise in the early universe}

\author{
    \href{https://orcid.org/0000-0003-4388-5606}{\includegraphics[scale=0.08]{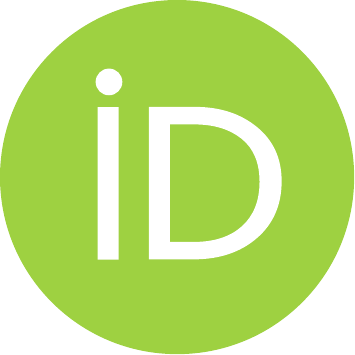}
    \hspace{1mm}Ezequiel Lozano} \\
    Departamento de Física \\ 
    Facultad de Ciencias Exactas y Naturales (FCEyN) \\ 
    Universidad de Buenos Aires (UBA) \\ 
    Ciudad Universitaria, 1428 Buenos Aires, Argentina. \\
    \texttt{elozano@df.uba.ar} \\
\And
    \href{https://orcid.org/0000-0001-7937-5419}{\includegraphics[scale=0.08]{orcid.pdf}
    \hspace{1mm}Francisco D. Mazzitelli} \\
    Centro Atómico Bariloche and Instituto Balseiro \\ 
    Comisión Nacional de Energía Atómica and CONICET \\ 
    Av E. Bustillo 9500, R8402AGP Bariloche, Argentina \\
    \texttt{fdmazzi@cab.cnea.gov.ar}
}

\renewcommand{\headeright}{The role of noise in the early universe}
\renewcommand{\undertitle}{}
\renewcommand{\shorttitle}{}

\hypersetup{
pdftitle={The role of noise in the early universe},
pdfsubject={gr-qc, hep-th},
pdfauthor={E.~Lozano, F.D.~Mazzitelli},
pdfkeywords={cosmological constant problem --- stochastic gravity --- Kapitza stabilization},
}

\begin{document}
\maketitle

\begin{abstract}
    We consider a quantum mechanical system to model the effect of quantum fields on the evolution of the early universe. The system consists of an inverted oscillator bilinearly coupled to a set of harmonic oscillators. We point out that the role of noise may be crucial in the dynamics of the oscillator, which is analyzed using the theory of harmonic oscillators with random frequency. Using this analogy we argue that, due to the fluctuations around its mean value, a positive vacuum energy density would not produce an exponentially expanding but an oscillating universe, in the same fashion that an inverted pendulum is stabilized by random oscillations of the suspension point (stochastic Kapitza pendulum). The results emphasize the relevance of noise in the evolution of the scale factor.
\end{abstract}


\section{Introduction}

The dynamical evolution equation for the scale factor in flat Robertson-Walker metrics reads
\begin{equation}
    \label{dyneq}
    \ddot{a} + \Omega^2 \, a = 0 \, , \, \quad \Omega^2 = \frac{4\pi \, G}{3}(\rho + 3 \, p) \, ,
\end{equation}
where $\rho$ and $p$ are the energy density and pressure that fill the universe. In the semiclassical version of this equation, $\rho$ and $p$ are replaced by the corresponding mean values of the energy density $\expval{\rho}$ and pressure $\expval{p}$ of the quantum fields.
If $\expval{\rho} = -\expval{p} > 0$, then $\expval{\Omega^2} = -8\pi \, G \expval{\rho}\!/ 3 $ is negative and generates an exponential evolution of the scale factor.

In the usual formulation of the cosmological constant problem (CCP) \cite{Weinberg, Martin, Padmanabham, Peebles}, the zero-point energy density $\expval{\rho}$ associated with a quantum field is assumed to be of the order of $E_{Pl}^{\,4}$, where $E_{Pl}$ is the Planck energy. This would produce a cosmological constant of order $E_{Pl}^{\,2}$, about $122$ orders of magnitude larger than the observed one. This result is obtained either on dimensional grounds or by considering a quantum field in flat spacetime, with an ultraviolet cutoff in $3-$momentum space $\Lambda_{cutoff}$ of the order of $E_{Pl}$. Strictly speaking, this procedure does not produce a cosmological constant, since evaluating the rest of the components of the energy momentum tensor one gets $\expval{p} = \expval{\rho}\!/3$, that is, the equation of state of a relativistic fluid \cite{Martin}. The physical reason behind this result is that the regularization method does not respect Lorentz invariance.

If the mean value of the stress tensor of the quantum field is covariantly regularized (for instance dimensional regularization with minimal subtraction) one obtains $\expval{T_{\mu\nu}} \propto m^4 \log(m / \mu) \, g_{\mu\nu}$, which corresponds to a cosmological constant of order $G \, m^4 \equiv m^4 / E_{Pl}^{\,2}$. Here $m$ is the mass of the quantum field and $\mu$ a renormalization scale. The effective cosmological constant in a given model depends on the particle content of the theory, and the choice of the renormalization scale. When considering the contributions of all fields in the Standard Model, with the corresponding sign for bosons and fermions, the effective cosmological constant turns out to be much smaller than $E_{Pl}^{\,2}$, but still $54$ orders of magnitude larger than the observed \cite{Martin}. We will denote the effective cosmological constant by $m_{eff}^{\,4} / E_{Pl}^{\,2}$, and will assume that $m_{eff}< E_{Pl}$. Other covariant methods of regularization, like the Pauli-Villars method or a $4$-momentum cutoff produce similar results, that is, $\expval{T_{\mu\nu}} \propto g_{\mu\nu}$ \cite{Ahkmedov, Ossola, Prokopec, Donoghue}. Depending on the details of the regularization, the main contribution to the cosmological constant is proportional to $m^4$, as in dimensional regularization, or to $m^2\Lambda_{cutoff}^2$. Note that in these approaches the contribution to the cosmological constant vanishes for massless fields, as has been recently emphasized in Ref.~\cite{Donoghue}.

The pressure and the energy density will have fluctuations around their mean values, and therefore we write $\rho = \expval{\rho} + \xi_\rho$, where the ``noise'' $\xi_\rho$ has a correlation function \cite{Verdaguer}
\begin{equation}
    N_\rho(t, t') = \frac{1}{2} \expval{\acomm{\rho(t) - \expval{\rho}\!(t)}{\rho(t') - \expval{\rho}\!(t')}} \, ,
\end{equation}
and similar expressions for the pressure.
The stochastic equation for the scale factor reads
\begin{equation}
    \label{see}
    \ddot{a} + \frac{4\pi \, G}{3}(\expval{\rho} + 3 \expval{p} + \xi_\rho + 3 \, \xi_p) \, a = 0 \, .
\end{equation}
In what follows we will be interested in a situation in which the mean values and their fluctuations correspond to the vacuum state, at scales where quantum fluctuations are relevant.   This could be the case in the very early universe, but still at sub-Planckian energies, in order to avoid quantum gravity effects. However, most of our discussion will be kept at a conceptual level, without reference to a particular cosmological epoch.

This is of course a toy model for the analysis of the effect of noise on the expansion of the universe. A more complete study should include spatial fluctuations of the energy momentum tensor, a general metric, and the consideration not only of the evolution equations but also of the constraints of General Relativity. These issues have been recently discussed in a series of papers by Wang et al. \cite{Wang, Wang2}. It was first pointed out that for metrics of the form
\begin{equation}\label{alocal}
    \dd{s}^2 = -\dd{t}^2 + a^2(t, \vec{x})(\dd{x}^2 + \dd{y}^2 + \dd{z}^2) \, ,
\end{equation}
the local scale factor $a(t, \vec{x})$ satisfies an equation similar to \cref{see}, at each spatial point $\vec{x}$. Assuming a negative cosmological constant, the scale factor would oscillate around $a = 0$ at the classical or semiclassical level. Including the noise term, these oscillations are parametrically amplified. This microscopically oscillating universe, when viewed at a macroscopic scale, would expand exponentially with a Hubble constant given by the parametric resonance, which could be much lower than expected if the resonance is weak \cite{Wang}. This happens when the bare cosmological constant is the largest scale in the model, even larger than the maximum frequency in the power spectrum of the noise kernel. A more general approach, that includes a discussion of the constraints of General Relativity, gives rise to a similar evolution equation \cite{Wang2}, while the constraints would imply that active fluctuations of the gravitational field can hide a large and negative cosmological constant. A similar mechanism has been described in Ref.~\cite{Carlip}, without assumptions about the sign of the cosmological constant. There is an ongoing debate on this \cite{CarlipvsWang}. There has been also discussions regarding some particular issues of the original proposal in Ref.~\cite{Wang} (see for instance Refs.~\cite{comments} and \cite{epjc}).

Despite the technical details, the main point raised in the analysis of Refs.~\cite{Wang,Wang2,Carlip} is that the CCP should be reformulated taking into account the inhomogeneities that come both from the active fluctuations of the metric, originated by quantum gravity, and from passive fluctuations induced by the quantum fields. The influence of stochastic variations of the Newton constant (originated by quantum geometry effects) on the late time expansion of the Universe has also been considered in Ref.~\cite{Fedele}.

In this paper we will focus on the effects of the quantum fields on the metric, when one takes into account not only the mean value of the energy momentum tensor as a source of the Einstein equations (semiclassical Einstein equations (SEE)) but also the fluctuations around this mean value (semiclassical stochastic gravity). Our main goal is to shed light on the following basic question: which is the role of the noise in the evolution of the local scale factor?

To avoid the technical complications of General Relativity, our discussion will be based on a quantum mechanical analog of the problem, which leads to an evolution equation similar to \cref{see}. Noise usually arises in the context of quantum open systems, and we will consider the paradigmatic example of quantum Brownian motion (QBM) \cite{HPZ1}, in which a heavy particle (the Brownian particle \cite{aclar}) is coupled to an environment. The effect of the environment is to modify the classical dynamics of the Brownian particle, which should now be described with a Langevin equation. Noise and dissipation  are the main consequences of the interaction with the environment.

Specifically, we will study a Brownian particle with an inverted oscillator potential, coupled to an environment composed by a set of harmonic oscillators. For a bilinear coupling, the dynamics of the Brownian particle are similar to that of the scale factor of the Universe, and the effects of the environment on the dynamics of the Brownian particle can be identified with those of the quantum fields.
As we will see, depending on the characteristics of the environment, the noise can change qualitatively the behavior of the Brownian particle. This analysis can be addressed using well known results of the theory of harmonic oscillators with multiplicative noise (or random frequency). When translated to the CCP, this means that there could be a ``noise induced stabilization'' and, instead of an exponential expansion, one could have rapid oscillations of the scale factor. Moreover, the toy model is useful to understand the validity (or not) of some of the approximations done in the gravitational problem.

\section{The model}

We will consider a Brownian particle with coordinate $x(t)$ coupled to a set of harmonic oscillators $q_n(t)$. The classical action is given by
\begin{equation}
    \label{classaction}
    S = \frac{1}{2} \int \dd{t} \left[ M\dot{x}^2 + M\Omega_B^{\,2} \, x^2 + \sum_n \left(m\dot{q}_n^{\,2} - m\omega_n^{\,2} \, q_n^{\,2} + g_n \, x^2 \, q_n^{\,2} \right) \right] \, .
\end{equation}
Note that the Brownian particle is an inverted harmonic oscillator, although later on we will also consider the case $\Omega_B^{\,2} < 0$.

In order to avoid unnecessary constants, in what follows we will make the replacements $\sqrt{M} \, x \to x$  and $\sqrt{m} \, q_n \to q_n$, so in this section $x$ and $q_n$ have units of [length]$^{1/2}$. We will also redefine the coupling constants as $\lambda_n = g_n / (M\,m)$.

The classical equations of motion read
\begin{align}
     & \ddot{x} - \left(\Omega_B^{\,2} + \sum_n \lambda_n \, q_n^{\,2}\right) x = 0 \, ,     \\
     & \ddot{q}_n + \left(\omega_n^{\,2} - \lambda_n \, x^2\right) q_n = 0 \, . \label{Heis}
\end{align}

In a semiclassical approximation, the oscillators in the environment are treated as quantum oscillators, and the Brownian particle classically. The semiclassical equation of motion for the Brownian particle is
\begin{equation}
    \label{SQBM}
    \ddot{x} - \left(\Omega_B^{\,2} + \sum_n \, \lambda_n \expval{q_n^{\,2}}\right) x = 0\, ,
\end{equation}
where the mean values $\expval{q_n^{\,2}}$ are computed in a pure or mixed state for the oscillators $q_n$. Note that \cref{Heis} is now the equation for the quantum operators in the Heisenberg picture. Taking into account the fluctuations around the mean values $\expval{q_n^{\,2}}$, one can derive a Langevin equation for the Brownian particle, that is of the form \cite{HPZ2}
\begin{equation}
    \label{LQBM}
    \ddot{x} - \left(\Omega_B^{\,2} + \sum_n \lambda_n \expval{q_n^{\,2}} + \xi(t)\right) x = 0 \, ,
\end{equation}
where
\begin{equation}
    \label{noise}
    \xi(t) = \sum_n \lambda_n \, \xi_n(t) \, , \, \quad \xi_n(t) = q_n^{\,2}(t) - \expval{q_n^{\,2}(t)} \, ,
\end{equation}
and $\xi_n$ are stochastic variables with  correlation functions
\begin{align}
    \label{RQBM}
    R_n(t - t') & = \frac{1}{2} \expval{\acomm{{\xi_n}(t)}{{\xi_n}(t')}} \, , \\
    R(t-t')     & = \sum_n \lambda_n^2 \, R_n(t - t') \, .
\end{align}

The mean values $\expval{q_n^{\,2}}$ and the correlation functions $R_n(t, t')$ are complicated functionals of the position of the Brownian particle $x(t)$.
The term $\sum_n \lambda_n \expval{q_n^{\,2}}$ renormalizes the bare frequency of the Brownian particle and introduces dissipation in the system.

We would like to use this theory as a toy model for the CCP. The coordinate of the Brownian particle plays the role of the scale factor of the universe, the environment that of the quantum fields, and $\Omega_B^{\,2}$ is the analog of the bare cosmological constant. \cref{SQBM} does not take into account the role of noise, and is the analog of the SEE in quantum field theory in curved spaces.
\cref{LQBM} is the analog of the Einstein-Langevin equations in stochastic gravity \cite{Verdaguer}.

When the effect of the Brownian particle on the environment is neglected, the $q_n$ become harmonic oscillators with frequency $\omega_n$. Considering thermal states of these oscillators at a temperature $\beta^{-1}$, we get
\begin{equation}
    \expval{q_n^{\,2}} =  \frac{Y(\omega_n)}{2 \, \omega_n} \, ,
\end{equation}
where  $Y(\omega_n) = \coth(\beta \, \omega_n / 2)$.  Thus the effect of the environment is to renormalize the ``bare cosmological constant'' $\Omega_B^{\,2}$
\begin{equation}
    \Omega_{ren}^{\,2} = \Omega_B^{\,2} + \sum_n \frac{\lambda_n \, Y(\omega_n)}{2 \, \omega_n} \, .
\end{equation}
Note that when $\Omega_{ren}^{\,2} > 0$, the semiclassical \cref{SQBM} implies exponential solutions for $x(t)$.
In the same approximation, the noise correlation functions read
\begin{equation}
    R_n(t - t') = \frac{1}{4\,\omega_n^{\,2}} \left( \cos(2\,\omega_n\,(t - t')) \left(Y(\omega_n)^{\,2} + 1\right) + Y(\omega_n)^{\,2} - 1 \right) \, .
\end{equation}
Neglecting the influence of the Brownian particle on the environment is the analog of ignoring the influence of the curved geometry in the evaluation of the mean value and fluctuations of the energy-momentum tensor when discussing the CCP.

The effect of the Brownian particle on the environment oscillators can be taken into account perturbatively in the coupling constants $\lambda_n$. Solving \cref{Heis} up to first order we obtain
\begin{equation}
    \label{eq:241}
    \expval{q_n^{\,2}} = \frac{Y(\omega_n)}{2 \, \omega_n} - \frac{Y(\omega_n) \, \lambda_n}{2\,\omega_n^{\,2}} \int^t \dd{t'} \sin(2 \, \omega_n \, (t - t')) \, x^2 \, .
\end{equation}
Inserting this result into \cref{LQBM} we get
\begin{equation}
    \label{eq:105}
    \ddot{x} + \Omega_{ren}^{\,2} \, x + x \int^t \dd{t'} D(t - t') \, x(t')^2 = {\xi} \, x \, ,
\end{equation}
where
\begin{equation}
    D(t - t') = -\sum_n \frac{Y(\omega_n) \, \lambda_n^2}{2 \, \omega_n^{\,2}} \, \sin(2 \, \omega_n(t - t')) \, , \label{eq:161}
\end{equation}
is the dissipation kernel.

Introducing the spectral density of the environment
\begin{equation}
    \label{eq:167}
    I(\omega) = \sum_n \, \delta(\omega - \omega_n) \, \frac{\lambda_n^2}{(2 \, \omega_n)^2} \, ,
\end{equation}
the noise and dissipation kernels can be written as
\begin{align}
    \label{eq:111}
     & R(t - t') = \int_{0}^\infty \dd{\omega} I(\omega) \left( (Y(\omega)^2 + 1) \, \cos(2 \, \omega \, (t - t')) + Y(\omega)^2 - 1 \right) , \\
     & D(t - t') = 2 \int_{0}^\infty \dd{\omega} I(\omega) \, Y(\omega) \, \sin(2\,\omega(t - t')) \, . \label{eq:248}
\end{align}
These equations have been derived previously using functional methods \cite{HPZ2}, for general nonlinear couplings between the system and the environment.

\section{Parametric resonance and noise-induced stability}

When the mean values and noises are evaluated neglecting the effect of the Brownian particle on the environment, the effective stochastic equation for the Brownian particle is that of an inverted harmonic oscillator with random frequency. It also describes a rigid pendulum around its unstable vertical position with a random vertical motion of its pivot.

It is a well-known and somewhat curious fact of classical mechanics that the rigid pendulum can be stabilized in the inverted position when the pivot oscillates vertically, if the frequency and amplitude of the oscillations are within certain intervals. This was pointed out at the beginning of the 20th century by A. Stephenson \cite{Steve}. Decades later, P. Kapitza provided a theoretical explanation of the device, now known as the ``Kapitza pendulum'' \cite{Kapitza}.
The oscillations of the pivot produce a torque whose average (over a temporal scale much shorter than the natural frequency of the pendulum) overwhelms gravity and the pendulum becomes stable \cite{Landau}.
An important difference between the Kapitza pendulum and our toy model is that we have an inverted pendulum with a {\it random} motion of the pivot. The ``stochastic Kapitza pendulum'' has also been widely investigated, and under certain circumstances one can have noise-induced stability \cite{Ibrahim}.

On the other hand, if we consider an ordinary harmonic oscillator (that corresponds to $\Omega_B^{\,2} < 0$ in our model), the stochastic equation is that of an oscillator with random frequency. For a harmonic oscillator with a deterministic time dependent frequency, there is parametric resonance if the time dependence contains non vanishing Fourier modes at $2 \, \omega_0 / n$, for  $n = 1,2,3,...$ \cite{Landau}. There are similar results when the frequency of the oscillator has a random component, as we will see below.

\subsection{Parametric resonance for a harmonic oscillator with random frequency}

Let us consider a harmonic oscillator with random frequency
\begin{equation}
    \label{eq:106}
    \ddot x + \Gamma \, \epsilon^2 \, \dot{x} + \omega_0^{\,2} \, (1 + \epsilon \, \xi(t)) \, x = 0 \, ,
\end{equation}
where  $\xi$ is a stochastic function with vanishing mean value and correlation function
\begin{equation}
    \label{RPR}
    S(t - t') = \expval{\xi(t) \, \xi(t')} \, .
\end{equation}
For further discussion, we added the dissipative term proportional to the positive constant $\Gamma$ and a small parameter $\epsilon$.

Using multiple scale analysis one can show that \cite{Pap}, to lowest order in $\epsilon$,
\begin{equation}
    \begin{aligned}
        \expval{x^2(t)} = \  & \frac{1}{2} \, \exp(\left(\frac{\omega_0^{\,2}}{2}\left(\Re{\tilde S(2 \, \omega_0)} - \tilde S(0)\right) - \Gamma\right)\epsilon^2 \, t) \, \times \\
                             & \times \cos(\left(2 \, \omega_0 - \frac{\epsilon^2\,\omega_0^{\,2}}{2}\,\Im{\tilde S(2 \, \omega_0)} \right) t) \, +                                \\
                             & + \frac{1}{2} \, \exp(\left( \omega_0^{\,2} \, \Re{\tilde S(2 \, \omega_0)} - \Gamma \right) \epsilon^2 \, t) + \order{\epsilon^2} \, ,
    \end{aligned}
\end{equation}
where $\tilde{S}(\omega)$ is the Fourier transform of the correlation function:
\begin{equation}
    \tilde{S}(\omega) = \int_0^\infty S(t) \, e^{i \, \omega \, t} \dd{t}\, . \label{eq:208}
\end{equation}
Therefore, there will be parametric resonance with a rate $\epsilon^2\,(\gamma - \Gamma)$ where
\begin{equation}
    \label{gammaRP}
    \gamma =  \omega_0^{\,2} \Re{\tilde{S}(2 \, \omega_0)} \, ,
\end{equation}
as long as $\gamma > \Gamma$.
In particular, for non dissipative systems parametric resonance takes place as long as $\Re{\tilde{S}(2 \, \omega_0)} > 0$. Working at higher orders in $\epsilon$, we expect a similar condition with the Fourier transform evaluated at $2 \, \omega_0 / n$ and a rate of growth proportional to $\epsilon^{n + 1}$, that tends to zero as $n \gg 1$. In this case we would have a ``weak'' parametric resonance.

When evaluated for the Brownian particle, setting $\epsilon = 1$ and neglecting the dissipative term, the condition for parametric resonance reads
\begin{equation}
    \Omega_{ren}^{\,2} \Re{\tilde{S}(2 \, \Omega_{ren})} = \frac{\pi}{4} \, \frac{I\!\left(\Omega_{ren}\right)}{\Omega_{ren}^{\,2}}\left(Y(\Omega_{ren})^2 + 1 \right) > 0 \, ,
\end{equation}
where we have taken into account that the definition of the noise in \cref{eq:106} differs from that of \cref{LQBM} by a factor $\omega_0^{\,2}$. In our QBM model, there will be parametric resonance if the spectral density of the environment satisfies $I(\Omega_{ren}) \neq 0$ and the dissipative term is negligible.

\subsection{Noise induced stability for an inverted harmonic oscillator with random frequency}

Let us now consider the linearized Kapitza pendulum
\begin{equation}
    \label{eq:112}
    \ddot{x} + \Gamma \, \dot{x} - \omega_0^{\,2} \left( 1 + \xi(t) \right) x = 0 \, ,
\end{equation}
where $x$ is the angle with respect to the vertical, $\omega_0^{\,2} = g / l$, $l$ is the length of the pendulum and $g$ the gravitational acceleration.
The stochastic function $\xi(t)$ describes the random acceleration of the pivot, $\xi(t) = \ddot{\varphi}(t)$, where $\varphi(t)$ is its position. We have also included a dissipative term.

The stability condition reads \cite{Howe}
\begin{equation}
    \label{cond1Howe}
    \sigma^2 \equiv \expval{\dot{\varphi}^2(t)} > \frac{1}{\omega_0^{\,2}} \, .
\end{equation}
If this condition is fulfilled the stochastic pendulum will be stable, and will oscillate around $a = 0$ with a frequency $\omega_K^{\,2} = \omega_0^{\,2}((\omega_0 \, \sigma)^2 - 1)$.

However, this is not the whole story. As we have seen, the ordinary harmonic oscillator with natural frequency $\omega_0$ can suffer parametric amplification when the Fourier transform of the noise kernel is non vanishing at $2 \, \omega_0$. The same happens for the ``stochastic Kapitza pendulum'': once stabilized, it still suffers a secular growth of the amplitude of oscillations when the Fourier transform of the noise correlation function is different from zero at $2 \, \omega_K$. The amplitude grows exponentially with a rate \cite{Howe} $\gamma_K - \Gamma$ where
\begin{equation}
    \gamma_K = \frac{\pi}{\omega_K^2} \left( 16\,\omega_K^4\,\Phi_{11}(2\,\omega_K) - 4\,\omega_K^2\left( \Phi_{12}(2\,\omega_K) + \Phi_{21}(2\,\omega_K) \right) + \Phi_{22}(2\,\omega_K) \right) ,
\end{equation}
and
\begin{align}
     & \Phi_{ij}(\omega) = \frac{1}{2\pi} \int_{-\infty}^\infty R_{ij}(t) \, e^{i \, \omega \, t} \dd{t} \, , \label{eq:119}                            \\
     & R_{ij}(t) = \expval{\eta_i(s) \, \eta_j(s + t)} \, , \label{eq:120}                                                                              \\
     & \eta_1(t) = \omega_0^{\,2} \, \varphi(t) \, , \label{eq:121}                                                                                     \\
     & \eta_2(t) = \omega_0^{\,4} \, \ddot{\varphi} (t) \, \varphi(t) - \omega_0^{\,4} \, \expval{\ddot{\varphi}(t)\ , \varphi(t)} \, .  \label{eq:122}
\end{align}

For our QBM model, the stability condition \eqref{cond1Howe} can be obtained by integrating twice the noise kernel in \cref{eq:248} with respect to $t$, and evaluating at $t = t'$. It reads
\begin{equation}
    \label{stab}
    \Omega_0^2 \equiv  \int_{0}^\infty \dd{\omega} \frac{I(\omega)}{4 \, \omega^2} \left(Y^2 + 1 \right) > \abs{\Omega_{ren}^{\,2}} \, ,
\end{equation}
and, when fulfilled, the oscillations have a frequency $\omega_K^{\,2} = \Omega_{ren}^{\,2} \left((\Omega_0 / \Omega_{ren})^2 - 1\right)$.

Note that the stability condition depends on the properties of the spectral function $I(\omega)$. It is usual to consider $I(\omega)\propto \omega^\alpha$, with $\alpha=1$ for ohmic environments and $\alpha>1$ for supraohmic environments. For these cases, it is necessary to introduce an ultraviolet frequency cutoff in the integral that defines $\Omega_0^{\,2}$. The cutoff corresponds to the maximum frequency of the oscillators in the environment, and the results will depend on this physical input.

To compute $\gamma_K$ we first note that it contains contributions from 2, 3 and 4-point correlation functions. From \cref{noise} we see that these are quadratic, cubic and quartic in the coupling constants $\lambda_n$, respectively. The leading contribution comes from $\Phi_{11}$ and is given by
\begin{equation}
    \label{gamma11}
    \gamma_K \simeq 8\pi \frac{I(2 \, \omega_K)}{\omega_K^{\,2}}(Y^2 + 1)\, .
\end{equation}

The general conclusion is the following. When analyzing the evolution of the Brownian particle at the semiclassical level, neglecting the effect of the noise, the evolution is unstable, and the solutions grow exponentially. This is the usual instability of the Kapitza pendulum. However, the noise can stabilize the system if the condition \cref{stab} is fulfilled. The Brownian particle will oscillate around $x = 0$ with a frequency $\omega_K^{\,2}$. There is an additional secular effect, that produces exponential growth in the amplitude of the oscillations, at a rate that is not given by the semiclassical equation, but is determined by the Fourier components of different noise kernels evaluated at $2 \, \omega_K$. The exponential growth takes place only in a small dissipation regime, i.e. when the term proportional to the dissipation kernel in \cref{eq:105} can be neglected. Note that this will be true as long as $x$ is sufficiently small.

\section{Application to the evolution of the scale factor of the universe}

With the caveats mentioned in the Introduction, we will now show that similar phenomena are produced by the fluctuations of the vacuum energy-momentum tensor around its mean value on the evolution of the scale factor.
Before doing this, we will discuss some properties of the noise kernel in quantum field theory.

\subsection{Fluctuations of the energy-momentum tensor}

The noise kernel (or correlation function) is formally defined as
\begin{equation}
    N_{\mu\nu\rho\sigma} = \frac{1}{2}\expval{\acomm{t_{\mu\nu}(x)}{t_{\rho\sigma}(x')}} \, ,
\end{equation}
where $t_{\mu\nu} = T_{\mu\nu} - \expval{T_{\mu\nu}}$.
The quantum fluctuations of the stress-tensor play a role in different physical phenomena like fluctuations of the Casimir and Casimir-Polder forces, radiation pressure fluctuations, black hole evaporation, etc. Although defined in terms of the unrenormalized stress-tensor, it is free from ultraviolet divergences, because the regularized and the renormalized stress-tensors differ by the identity operator times geometric counterterms, and then $t_{\mu\nu}$ is equal to its renormalized counterpart \cite{Verdaguer}.
This has been explicitly checked in Ref.~\cite{FordWoodard}, in the context of the Schwinger-Keldysh formalism. Working in $d$ dimensions, it was shown there that while for $d\to 4$ the real part of the effective action contain the divergences that must be absorbed into the bare constants of the theory, the imaginary part, that determines the noise kernel, is a well defined distribution, and no divergences show up.

The noise kernel does have a divergence in the coincidence limit $x \to x'$, which is of a different nature.
Physical observables involve integrals of the correlation function. In particular, the smeared correlation function describes the finite fluctuations of the average of the stress-tensor over a finite spacetime region \cite{Fewster}.
For instance, the fluctuations of Casimir force have been obtained by averaging the correlation function over the time scale associated with the measurement process \cite{Barton}. Gravitational consequences of the noise have also been considered in different situations \cite{Fords}, with the corresponding smearing of the correlation function. Although there have been some attempts to renormalize the correlation function using point-splitting regularization combined with Hadamard subtraction \cite{Hu2001}, the physical interpretation of the renormalized coincidence limit of the noise kernel is not clear, and the smearing of the correlation function was considered in later works \cite{later}.
We will take here a similar point of view. We stress that the divergence in the coincidence limit is similar to the one in QBM mentioned after Eq.\eqref{stab}, and could be treated accordingly, by introducing a cutoff in the modes of the quantum field (the introduction of a cutoff corresponds to a smearing with a particular weight function).

\subsection{Positive cosmological constant: Kapitza stabilization}

We now compare the evolution equation of the scale factor with that of the linearized Kapitza pendulum.
We stress once again that we are not taking into account the spatial fluctuations of the energy-momentum tensor, that would produce spatial inhomogeneities in the metric. Hopefully, a more rigorous approach can be applied to the evolution of the local scale factor,
as described below in Section \ref{constraints}.

Evaluating the mean values and noises in Minkowski spacetime, from \cref{see} we obtain
\begin{equation}
    \label{seeMink}
    \ddot{a} -\frac{8\pi \, G}{3} \left(\expval{\rho} + \xi_\rho\right) a = 0\, ,
\end{equation}
where we assumed a Lorentz invariant regularization, i.e. $\expval{p} = -\expval{\rho}$.
Comparing this equation with \cref{eq:112} we make the identifications
\begin{align}
    \omega_0^{\,2} & \to 8\pi \, G \expval{\rho} / 3 \, ,      \\
    \xi(t)         & \to \frac{\xi_\rho(t)}{\expval{\rho}}\, .
\end{align}
Introducing a new stochastic variable $\varphi_{\!\rho}(t)$ such that $\ddot{\varphi}_{\!\rho}(t) = \xi_\rho(t)$ the stability condition reads
\begin{equation}\label{eq:stab}
    \expval{\dot{\varphi}_{\!\rho}^{\,2}(t)} > \frac{3 \expval{\rho}}{8\pi \, G} \, .
\end{equation}

As already mentioned, the correlation function is a distribution that diverges in the coincidence limit.
In fact, for a massless field one can easily check that
\begin{equation}
    \label{Nrho}
    N_{\!\rho}(t,t') \propto \frac{1}{(t - t')^8} \, ,
\end{equation}
a result that is valid even for massive fields when $t \to t'$. This implies that
\begin{equation}
    \expval{\dot{\varphi}_{\!\rho}(t) \, \dot{\varphi}_{\!\rho}(t')} \propto \frac{1}{(t - t')^6} \, .
\end{equation}
As discussed in the previous subsection, a smearing is unavoidable in order to obtain physical results. The scale of smearing could be related to the discreteness of spacetime in a quantum theory of gravity. If the correlation is evaluated for quantum fields with a smearing over a finite spacetime region, on dimensional grounds we expect
\begin{equation}
    \label{fluct}
    \expval{\dot{\varphi}_{\!\rho}^{\,2}(t)} \propto N_{\!f} \, E_0^{\,6} \, ,
\end{equation}
where $E_0$ is the scale of smearing and $N_{\!f}$ the number of fields.

Given the relevance of this result for our discussion, we illustrate it with a simple example: the two-point function of a massless scalar field $G(t,{\mathbf{x}},t',{\mathbf{x}}) = \expval{\acomm{\phi(t, \mathbf{x})}{\phi(t', \mathbf{x})}}$ diverges as $(t - t')^{-2}$ in the coincidence limit. However, for fields with a Gaussian smearing over a time scale $E_0^{\,-1}$ we have \cite{HuPhillips}
\begin{equation}
    \bar{G}(t, {\mathbf{x}}, t', {\mathbf{x}}) = \frac{E_0^{\,2}}{8\pi^2} \left(1 - \frac{(t - t')^2}{4} \, E_0^{\,2}\right) ,
\end{equation}
where a bar denotes the smearing. Note that, in the coincidence limit $\bar{G} \simeq E_0^{\,2}$, as expected on dimensional grounds. Moreover
\begin{equation}
    \label{Gbar}
    \expval{\acomm{\pdv[2]{t} \bar{\phi}(t, \mathbf{x})}{\bar{\phi}(t, \mathbf{x})}} = -\frac{1}{16\pi^2} \, E_0^{\,4} \, .
\end{equation}
A similar argument leads to \cref{fluct}, which is valid even for massive fields, as long as the mass of the field is much smaller than $E_0$.

Writing $E_0 = \chi \, E_{Pl}$, the stability condition reads $N_f \,\chi^6 \, E_{Pl}^{\,4} > m_{eff}^{\,4}$, which is easily fulfilled in the Standard Model for $\chi = \order{1}$. Therefore, the noise changes drastically the behavior of the scale factor, which would suffer oscillations around $a = 0$ with a frequency
\begin{equation}
    \omega_K^{\,2} = N_{\!f} \, \chi^6 \, E_{Pl}^{\,2} - m_{eff}^{\,4} / E_{Pl}^{\,2} \simeq N_{\!f} \, \chi^6 \, E_{Pl}^{\,2} \, .
\end{equation}
The singularities of the geometry produced when $a = 0$ will not be discussed here, since we are only pointing out the qualitative difference between the expected behavior of the scale factor with and without noise.

Regarding parametric amplification, using \cref{gamma11} and our dictionary to go from the stochastic pendulum to the evolution of the scale factor, we obtain
\begin{equation}
    \gamma_K = 16\pi \, \omega_K^{\,2} \left(\frac{8\pi \, G}{3}\right)^{\!2} \widetilde{Q}(2 \, \omega_K) \, ,
\end{equation}
where $\widetilde{Q}(2 \, \omega_K)$ is the Fourier transform of the correlation function $Q(t, t') = \expval{\varphi_{\!\rho}(t) \, \varphi_{\!\rho}(t')}$.
Taking into account that $Q(t, t')$ diverges as $(t - t')^{-4}$ we have
\begin{equation}
    \widetilde{Q}(2 \, \omega_k) = N_{\!f} \, \omega_K^{\,3} \, f(\omega_K / E_0) \, ,
\end{equation}
where $f(z)$ depends on the smearing and tends to a nonvanishing constant as $z \to 0$. Thus, there is an additional effect of the noise, which is to amplify the oscillations with a rate
\begin{equation}
    \gamma_K \propto N_{\!f} \, \frac{ \omega_K^{\,5}}{E_{Pl}^{\,4}} \, f\!\left(\omega_K / E_0\right) = N_{\!f}^{\,7/2} \, \chi^{15} \, E_{Pl} \, f(\sqrt{ N_{\!f}} \, \chi^2) \, .
\end{equation}

We would like to stress that, although we used dimensional regularization to evaluate $\expval{\rho}$, this is not crucial. The stabilization will work as long as \cref{eq:stab} is satisfied, independently of the regularization method.

\subsection{Negative cosmological constant: parametric resonance}

For the sake of completeness, we perform a similar analysis for the case of a negative cosmological constant. In the absence of noise, \cref{dyneq} describes a harmonic oscillator with frequency $\omega_0^{\,2} = 8\pi \, G \, \abs{\expval{\rho}}/3$, which is of order $m_{eff}^{\,4} / E_{Pl}^{\,2}$. Including the noise, the dynamical equation is that of a harmonic oscillator with a random frequency $\omega^2(t) = \omega_0^{\,2} + 8\pi \, G  \, \xi_\rho(t)/3$. According to the theory of parametric resonance for the usual harmonic oscillator with random frequency, reviewed in the previous section, the amplitude of the oscillations will grow exponentially with a rate given by \cref{gammaRP}
\begin{equation}
    \gamma = \omega_0^{\,2} \, \frac{\Re{\tilde{N}(2 \, \omega_0)}}{\expval{\rho}^2} \, .
\end{equation}
Taking into account \cref{Nrho}, the Fourier transform of the noise kernel is
\begin{equation}
    \widetilde{N}_{\!\rho}(2 \, \omega_0) = N_{\!f} \, \omega_0^7 \, f(\omega_0 / E_0) \, ,
\end{equation}
where, once more, the function $f(z)$ depends on the smearing.
These results imply that $\widetilde{N}_{\!\rho}(2 \, \omega_0)$ will be nonvanishing, since we are assuming that $\omega_0 \leq E_0$. Therefore, we expect a strong parametric resonance with
\begin{equation}
    \gamma \propto N_{\!f} \, \frac{ \omega_0^{\,5}}{E_{Pl}^{\,4}} \, f(\omega_0 / E_0) \, .
\end{equation}

Note that the scenario here is different from that of Ref.~\cite{Wang2}, where it is assumed that $\omega_0 \gg E_0$. In that case, the parametric resonance is weak, and the associate Hubble constant tends to zero as $\omega_0 \to \infty$.

\subsection{Spatial fluctuations and constraint equations}\label{constraints}

During this work, to maintain a close analogy with QBM, we deliberately omitted spatial fluctuations of the metric, and considered only the dynamical equation for the global scale factor $a(t)$. One can improve the description by considering the more general metric in Eq.\eqref{alocal}. Even if this is not the most general metric needed for arbitrary fluctuations of the quantum fields, its consideration sheds light on some important aspects of the CCP. We will follow closely Ref.~\cite{Wang}. As shown there, the fluctuations of the $0 j$-components of the stress tensor imply that the local Hubble constant defined by
\begin{equation}
    H(t, \vec{x}) = \frac{\dot{a}}{a}(t, \vec{x}) \, ,
\end{equation}
must differ from point to point. Moreover, the fluctuations of $T_{ij}$ imply that  the spatial derivatives of $a(t,\vec{x})$ must also be wildly fluctuating. The $00$-equation introduces a constraint between temporal and spatial fluctuations of the local scale factor.

On the other hand, the classical dynamical equation is analogous to
Eq.\eqref{dyneq}, that is
\begin{equation}\label{dyneqlocal}
    \ddot{a}(t, \vec{x}) + \Omega^2(t, \vec{x}) \, a(t, \vec{x}) = 0 \, ,
\end{equation}
where
\begin{equation}
    \Omega^2(t, \vec{x}) = T_{00} + \frac{1}{a^2}\sum_{i = 1}^3 T_{ii} \, .
\end{equation}
A similar equation holds for the local scale factor even for more general metrics \cite{Wang2,Carlip}, with an additional shear term.

In this context, an averaged Hubble rate can be defined as \cite{Wang}
\begin{equation}
    H(t)= \frac{\dot L}{L} \, ,
\end{equation}
where $L$ is the physical distance between points at different
spatial coordinates
\begin{equation}
    L(t) = \int_{\vec{x}_1}^{\vec{x}_2} \dd{l} \sqrt{a^2(t, \vec{x})} \, .
\end{equation}

Although the constraint equations have been treated at a qualitative level, these considerations show that spatial fluctuations are unavoidable. More importantly for our work, we stress that, when the mean value and fluctuations of $\Omega^2(t,\vec x)$ are evaluated using a quantum field in Minkowski spacetime, they are independent of the scale factor, and therefore the spatial coordinates enter parametrically in the dynamical equation. As both the stability condition and the condition for parametric resonance involve the coincidence limit or the Fourier transform of the correlation function at different times, both conditions will be $\vec{x}$-independent. The spatial dependence will be present only in the initial conditions needed to solve the dynamical equation.

\section{Outlook}

To summarize, the main implication  suggested by this analysis is as follows. When noise is neglected, and in the presence of a huge and positive cosmological constant, quantum matter fields in the vacuum state would produce an exponential expansion of the scale factor with a rate many orders of magnitude larger than observed.
However, as for the stochastic Kapitza pendulum, the noise generates rapid oscillations of the scale factor with a frequency $\sqrt{N_{\!f}} \, \chi^3 \, E_{Pl}$. Physically, the stabilization is produced because the fluctuations of the energy momentum tensor around its mean value are determined by the scale $\chi \, E_{Pl}$, which may be much larger than $m_{eff}$.
This is not the only consequence of the noise. As matter fields have a wide-band noise spectrum, the Fourier transform of the noise evaluated at twice the oscillation frequency will be non-vanishing, producing a secular amplification of the oscillations by parametric resonance. Depending on the parameters and matter content, the secular rate of expansion could eventually be smaller than $m_{eff}^{\,4} / E_{Pl}^{\,2}$.
In tune with the idea put forward in Refs.~\cite{Wang,Wang2}, we have shown, using an analog toy model, that the role of noise is crucial and has been overlooked in the formulation of the CCP. Interestingly, the analogy with the Kapitza pendulum has also been useful to analyze a complementary effect, in which stochastic fluctuations of the metric influence the dynamics of a classical scalar field during a phase transition \cite{Krukov}.

We have also considered the case of a negative cosmological constant. We pointed out that, when the mean values and noise kernels are evaluated using standard techniques, the noise produces a strong parametric resonance. This differs from the results of Ref.~\cite{Wang2}, since those works assume that the bare cosmological constant is much larger than the maximum frequency that is present in the noise kernel, producing a weak parametric resonance.

In all our analysis we have evaluated the mean value of the energy density and the noise kernel in Minkowski spacetime. This is of course a crude approximation that should be improved.
The QBM analog described in this paper illustrates this point. When the effect of the Brownian particle on the environment is neglected, the environment produces a renormalization of the frequency along with a noise term in the Langevin equation for the particle. However, when taken into account, it also produces a dissipative term. This is a general feature of the fluctuation-dissipation relation. We have seen that the traditional dissipative term, proportional to $\Gamma \, \dot{x}$, is relevant when discussing parametric resonance. A generalized dissipative term, like the one in our \cref{eq:105}, could also be relevant.
For quantum fields in curved spaces, when evaluating the mean value of the energy-momentum tensor beyond the Minkowski approximation, we expect the presence of generalized dissipative terms in the evolution of the local scale factor. This issue certainly deserves further study,  and it is crucial to understand the transition between a situation in which the noise is dominant (as described in this paper) and the usual description
of the expansion of the universe using a classical or semiclassical source in the Einstein equation.

Finally, we would like to point out that, although many aspects of this work are motivated by the approach of Ref.~\cite{Verdaguer}, a full understanding of the role of noise on the background metric needs an extension of the stochastic gravity program beyond the situations considered there, in which the noise is a small correction to the SEE.
This is an interesting avenue for future research, in which the analogy discussed here could also be useful.

\section*{Acknowledgments}

FDM is a member of CONICET. This research was supported by ANPCyT, CONICET, and UNCuyo.


\begin{thebibliography}{bib}
    \bibitem{Weinberg} S.~Weinberg,
    Rev. Mod. Phys. \textbf{61}, 1-23 (1989).
    \bibitem{Martin} J.~Martin,
    Comptes Rendus Physique {\bf 13}, 566 (2012).
    \bibitem{Padmanabham} T.~Padmanabhan,
    Phys. Rept. \textbf{380}, 235-320 (2003).
    \bibitem{Peebles} P.~J.~E.~Peebles and B.~Ratra,
    Rev. Mod. Phys. \textbf{75}, 559-606 (2003).
    \bibitem{Verdaguer} B.~L.~Hu and E.~Verdaguer, 
    Living Rev. Relativ. {\bf 11}, 3 (2008); {\it Semiclassical and Stochastic Gravity:
    Quantum Field Effects on Curved Spacetime}, Cambridge Monographs on Mathematical Physics, Cambridge University Press, Cambridge 2020.
    \bibitem{Ahkmedov} E.~K.~Akhmedov,
    [arXiv:hep-th/0204048 [hep-th]].
    \bibitem{Ossola} G.~Ossola and A.~Sirlin,
    Eur. Phys. J. C \textbf{31}, 165-175 (2003).
    \bibitem{Prokopec} J.~F.~Koksma and T.~Prokopec,
    [arXiv:1105.6296 [gr-qc]].
    \bibitem{Donoghue} J.~F.~Donoghue,
    [arXiv:2009.00728 [hep-th]].
    \bibitem{Wang} Q.~Wang, Z.~Zhu and W.~G.~Unruh,
    Phys. Rev. D \textbf{95}, 103504 (2017); 
    S.~S.~Cree, T.~M.~Davis, T.~C.~Ralph, Q.~Wang, Z.~Zhu and W.~G.~Unruh,
    Phys. Rev. D \textbf{98}, 063506 (2018).
    \bibitem{Wang2} Q.~Wang,
    Phys. Rev. Lett. \textbf{125}, 051301 (2020);
    Q.~Wang and W.~G.~Unruh,
    Phys. Rev. D \textbf{102}, 023537 (2020).
    \bibitem{Carlip} S.~Carlip,
    Phys. Rev. Lett. \textbf{123}, 131302 (2019).
    \bibitem{CarlipvsWang} Q.~Wang and W.~G.~Unruh,
    [arXiv:1911.06110 [gr-qc]]; 
    Q.~Wang and W.~G.~Unruh,
    Phys. Rev. Lett. \textbf{125}, 089001 (2020);
    S.~Carlip,
    [arXiv:1911.11203 [gr-qc]];
    S.~Carlip,
    Phys. Rev. Lett. \textbf{125}, 089002 (2020);
    O.~F.~Piattella,
    [arXiv:2007.02637 [gr-qc]].
    \bibitem{comments} F.~D.~Mazzitelli and L.~G.~Trombetta,
    Phys. Rev. D \textbf{97}, 068301 (2018); 
    Q.~Wang and W.~G.~Unruh,
    Phys. Rev. D \textbf{97},  068302 (2018).
    \bibitem{epjc} G.~R.~Bengochea, G.~León, E.~Okon and D.~Sudarsky,
    Eur. Phys. J. C \textbf{80}, 18 (2020).
    \bibitem{Fedele} M.~de Cesare, F.~Lizzi and M.~Sakellariadou,
    Phys. Lett. B \textbf{760}, 498 (2016).
    \bibitem{HPZ1}
    B.~L.~Hu, J.~P.~Paz and Y.~h.~Zhang,
    Phys. Rev. D \textbf{45}, 2843-2861 (1992).
    \bibitem{aclar} Please note that we are not using the term "Brownian particle" in the sense of the original Brownian motion, in which the mean squared displacement is proportional to $t$.
    \bibitem{HPZ2} B.~L.~Hu, J.~P.~Paz and Y.~Zhang,
    Phys. Rev. D \textbf{47}, 1576 (1993).
    \bibitem{Steve} A.~Stephenson, 
    Phil. Mag. {\bf 15}, 233 (1908).
    \bibitem{Kapitza} P.~L.~Kapitza, 
    {\it Dynamical stability of a pendulum when its point of suspension vibrates}, in {\it Collected papers of P. L. Kapitza, Volume 2}, D.~Ter~Haar (editor), Pergamon Press (1965).
    \bibitem{Landau} L.~D.~Landau and E.~M.~Lifshitz, 
    {\it Course of Theoretical Physics, Volume 1, Mechanics}, Pergamon Press, Oxford, 1969 (Second edition).
    \bibitem{Ibrahim} R.~A.~Ibrahim, 
    J. Vib. Control {\bf 12}, 1093 (2006).
    \bibitem{Pap} G.~Papanicolau and J.~B.~Keller, 
    SIAM J. Appl. Math. {\bf 21(2)}, 287 (1971).
    \bibitem{Howe} M.~S.~Howe, 
    J. Sound. Vib. {\bf 32}, 407 (1974).
    \bibitem{FordWoodard} L.~H.~Ford and R.~P.~Woodard,
    Class. Quant. Grav. \textbf{22}, 1637-1647 (2005).
    \bibitem{Fewster} C.~J.~Fewster and L.~H.~Ford,
    Phys. Rev. D \textbf{101}, 025006 (2020), and references therein.
    \bibitem{Barton} G.~Barton,
    J. Phys. A \textbf{24}, 991-1006 (1991).
    \bibitem{Fords} L.~H.~Ford and T.~A.~Roman,
    Phys. Rev. D \textbf{72}, 105010 (2005).
    \bibitem{Hu2001} N.~G.~Phillips and B.~L.~Hu,
    Phys. Rev. D \textbf{63}, 104001 (2001).
    \bibitem{later} B.~L.~Hu and A.~Roura,
    Phys. Rev. D \textbf{76}, 124018 (2007);
    A.~Eftekharzadeh, J.~D.~Bates, A.~Roura, P.~R.~Anderson and B.~L.~Hu,
    Phys. Rev. D \textbf{85}, 044037 (2012).
    \bibitem{HuPhillips} B.~L.~Hu and N.~G.~Phillips, 
    Phys. Rev. D {\bf 62}, 084017 (2000).
    \bibitem{Krukov} M.~Kurkov,
    Eur. Phys. J. C \textbf{76}, 329 (2016).
\end{thebibliography}
\end{document}